\documentclass[a4paper,12pt]{article}

\usepackage{amsmath}
\usepackage{amssymb}
\usepackage[dvips]{graphicx}
\usepackage[dvips]{psfrag}
\usepackage{color}

\makeatletter
\@addtoreset{equation}{section}
\renewcommand{\theequation}{\thesection.\@arabic\c@equation}
\makeatother

\makeatletter
\renewcommand\appendix{\par
  \setcounter{section}{0}%
  \setcounter{subsection}{0}%
  \gdef\thesection{Appendix \@Alph\c@section }
  \renewcommand{\theequation}
  {\Alph{section}.\arabic{equation}}
}
\makeatother

\newcommand{\dd}{{\rm d}}
\newcommand{\ee}{{\rm e}}

\begin{document}

\titlepage

\vspace*{-15mm}   
\baselineskip 10pt   
\begin{flushright}   
\begin{tabular}{r}    
{\tt APCTP-Pre2010-004}\\
{\tt HRI/ST/xxx}\\
{\tt YITP-10-63}\\
July 2010
\end{tabular}   
\end{flushright}   
\baselineskip 24pt   
\vglue 10mm   

\begin{center}
{\Large\bf
Notes on the Hidden Conformal Symmetry in \\
the Near Horizon Geometry of the Kerr Black Hole
}

\vspace{8mm}   

\baselineskip 18pt   

\renewcommand{\thefootnote}{\fnsymbol{footnote}}

Yoshinori Matsuo$^a$\footnote[2]{ymatsuo@hri.res.in}, 
Takuya Tsukioka$^b$\footnote[3]{tsukioka@apctp.org} 
and 
Chul-Moon Yoo$^c$\footnote[4]{yoo@yukawa.kyoto-u.ac.jp}

\renewcommand{\thefootnote}{\arabic{footnote}}
 
\vspace{5mm}   

{\it  
${}^a$ Harish-Chandra Research Institute, 
 Chhatnag Road, Jhusi, Allahabad 211 019, India \\
${}^b$ Asia Pacific Center for Theoretical Physics, 
 Pohang, Gyeongbuk 790-784, Korea \\
${}^c$ Yukawa Institute of Theoretical Physics, 
 Kyoto University, Kyoto 606-8502, Japan
}
  
\vspace{10mm}   

\end{center}

\begin{abstract}
Toward the Kerr/CFT correspondence for generic non-extremal Kerr 
black hole, 
the analysis of scattering amplitudes by near extremal Kerr 
provides a clue.     
This pursuit reveals a hidden conformal symmetry in 
the law frequency wave equation for a scalar field in a 
certain spacetime region referred to as the near region.   
For extremal case, the near region is expected to be the near 
horizon region in which the correspondence via the asymptotic 
symmetry is studied. 
We investigate the hidden conformal symmetry 
in the near horizon limit and consider the relation 
between the hidden conformal symmetry and the asymptotic 
symmetry in the near horizon limit. 
By using an appropriate definition of the quasi-local charge, 
we obtain the deviation of the entropy from the extremality. 
\end{abstract}

\baselineskip 18pt   

\newpage

\section{Introduction}\label{sec:Intro}

Recently, the correspondence between the Kerr black hole and 
conformal field theories was conjectured. 
The correspondence was studied in two different ways. 
The original correspondence was proposed in the near 
horizon geometry of extremal Kerr~\cite{ghss} 
whose isometry was $SL(2,\mathbb R)\times U(1)$~\cite{baho}. 
In \cite{ghss} they considered the enhancement of the $U(1)$ isometry 
to the Virasoro algebra by using the asymptotic symmetry \cite{bh}. 
This analysis is generalized to many other cases \cite{many}. 
After these works, an another Virasoro algebra was obtained 
by extending the $SL(2, \mathbb R)$ part of the isometry \cite{mty,cl}. 
The former Virasoro algebra is referred to as 
that of the left movers while the latter as that of the right movers. 

More recently, one observation toward generic Kerr/CFT correspondence 
was given in \cite{cms} after the works in which certain correlators 
in the dual CFT were obtained from the scattering amplitudes 
in the near extremal Kerr~\cite{bhss}. 
In the analysis for the scalar amplitude in \cite{cms}, 
an essential departure from the original Kerr/CFT correspondence 
is that it is not necessary to take the near horizon geometry. 
As in the case of AdS, the scalar wave equation can be solved 
by using the matching method at sufficiently low frequencies. 
In the analysis, first of all, 
the geometry is divided into two regions called as the near and 
the far region. 
These two regions have a overlap and 
the matching surface can be put at an arbitrary position 
in the overlap region. 
This arbitrariness implies the presence of the symmetry. 
In fact, the wave equation has the 
$SL(2,\mathbb R)_R\times SL(2,\mathbb R)_L$ symmetry 
under a suitable approximation. 
This symmetry is called as the hidden conformal symmetry. 
This implies that the scalar excitations in 
the Kerr geometry correspond to CFT. 
Applications of this analysis are studied in \cite{many2}. 

In this paper, we investigate relations between these two analysis. 
It can be expected that 
for the extremal case the near region 
reduces to the near horizon region 
and the conformal symmetry of the wave equation 
is connected to the symmetry of the geometry. 
We consider the equation of motion for scalar 
in the near horizon limit. 
We observe the relation between 
the $SL(2,\mathbb R)_R\times SL(2,\mathbb R)_L$ vectors 
and the asymptotic Killing vectors in the near horizon geometry. 
We extend the hidden conformal symmetry to the Virasoro algebra 
assuming that the extended vectors have 
the same form to those in AdS$_3$. 
We can consider the central charge by using 
the covariant definition of the asymptotic charge. 
However the result contains an extra term which should not appear. 
We can also access to the central charge by using appropriate 
definitions of the quasi-local charge. 
We could grub the origin of this peculiar term in the asymptotic charge  
and exclude it through a suitable definition. 
For the entropy, by using an another definition of 
the quasi-local charge which was given in our previous work~\cite{mty}, 
we observe that the Cardy formula reproduces 
the deviation of the entropy from the extremality. 

This paper is organized as follows:  
In section~\ref{sec:Kerr}, 
after providing the basic properties of the Kerr black hole, we briefly 
review on the hidden conformal symmetry. 
In Section~\ref{sec:NH}, we consider the near extremal case 
and introduce the near horizon limit. 
The extension of the vectors to the Virasoro generators 
is given with comparing the asymptotic Killing vector 
in Section~\ref{sec:Virasoro}.
We then estimate the central charge by using the asymptotic charge 
and also certain definitions of the quasi-local charge 
in Section~\ref{sec:Central}.  
In Section~\ref{sec:Entropy}, 
making use of the another definition of the quasi-local charge,  
we show the correspondence of the entropy through the Cardy formula. 
Section~\ref{sec:Disc} is devoted to discussions. 
One appendix is prepared.

\section{Kerr black hole and hidden conformal symmetry}\label{sec:Kerr}

We start by introducing the Kerr metric in Boyer-Lindquist coordinates: 
\begin{align}
 \dd s^2 &= -\dd t^2 
  + \frac{2Mr}{r^2 + a^2 \cos^2\theta} 
  \left(\dd t-a\sin^2\theta \dd\phi\right)^2 
  + \left(r^2 +a^2\right) \sin^2\theta \dd\phi^2 
 \notag\\&\quad
  + \frac{r^2 + a^2 \cos^2\theta}{r^2 -2Mr +a^2} \dd r^2 
  + \left(r^2 + a^2\cos^2\theta\right)\dd\theta^2. 
\end{align}
The parameters $M$ and $a$ are related to the ADM mass 
and the angular momentum as 
\begin{align}
 M_{\text{ADM}} &= \frac{M}{G_N} , & 
 J &= \frac{aM}{G_N} . 
\end{align}
The position of the horizon and the Hawking temperature are given by 
\begin{align}
 r_{\pm} &= M \pm \sqrt{M^2 -a^2} , &
 T_H &= \frac{r_+ - M}{4\pi M r_+} . 
\end{align}
The Bekenstein-Hawking entropy is 
\begin{equation}
S_{\rm BH}=\frac{\mbox{Area}}{4G_N}=\frac{2\pi Mr_+}{G_N}.
\label{bhe} 
\end{equation}

Let us now consider the Klein-Gordon equation for a massless scalar:
\begin{equation}
 \partial_\mu \left(
 \sqrt{-g} g^{\mu\nu} \partial_\nu \Phi(t,r,\phi,\theta)
 \right) = 0 . 
\end{equation}
By factorizing the scalar field as 
\begin{equation}
 \Phi(t,r,\phi,\theta) = \ee^{-i\omega t + i m \phi} R(r) S(\theta) , 
\end{equation}
this equation of motion can be separated into 
the following two differential equations with a separation constant $K$:  
\begin{subequations}
\begin{eqnarray}
&& 
\Bigl[
  \partial_r \Delta \partial_r 
  + \frac{(2Mr_+\omega-am)^2}{(r-r_+)(r_+-r_-)} 
  - \frac{(2Mr_-\omega-am)^2}{(r-r_-)(r_+-r_-)} 
 \notag
\nonumber 
\\
&&\quad
  + \Big(r^2 + 2 M(r+2M)\Big)\omega^2
 \Bigr] R(r) 
= K R(r) , 
 \label{RadialEqOriginal}
\end{eqnarray}
and 
\begin{equation}
 \left[
  \frac{1}{\sin\theta}\partial_\theta(\sin\theta\partial_\theta) 
  - \frac{m^2}{\sin^2\theta} + \omega^2 a^2 \cos^2\theta
 \right] S(\theta) = -K S(\theta) ,  \label{AngularEqOriginal}
\end{equation}
\end{subequations}
where $\Delta = (r-r_+)(r-r_-)$. 
One can consider the scalar excitation which has 
a large wavelength compared to the radius of curvature 
\begin{equation}
 \omega M \ll 1 , 
\end{equation}
and study its behavior in the near region, which is defined by 
\begin{equation}
 r \ll \frac{1}{\omega} . 
\end{equation}
Then, \eqref{RadialEqOriginal} and \eqref{AngularEqOriginal} 
can be approximated as 
\begin{subequations}
\begin{equation}
 \left[
  \partial_r \Delta \partial_r 
  + \frac{(2Mr_+\omega-am)^2}{(r-r_+)(r_+-r_-)} 
  - \frac{(2Mr_-\omega-am)^2}{(r-r_-)(r_+-r_-)} 
 \right] R(r) = K R(r) , \label{RadialEq}
\end{equation}
and 
\begin{equation}
 \left[
  \frac{1}{\sin\theta}\partial_\theta(\sin\theta\partial_\theta) 
  - \frac{m^2}{\sin^2\theta} 
 \right] S(\theta) = -K S(\theta) ,  \label{AngularEq}
\end{equation}
\end{subequations}
respectively. 
From \eqref{AngularEq}, one could obtain the separation constant  
\begin{equation}
 K = l(l+1) . 
\end{equation}

The equation \eqref{RadialEq} is solved by hypergeometric functions, 
which are representations of $SL(2,\mathbb R)$. 
To see this, it might be convenient to introduce 
the ``conformal'' coordinates $(w^+, w^-, y)$ which is defined by 
\begin{align}
 w^+ &= \sqrt{\frac{r - r_+}{r - r_-}} \ \ee^{2 \pi T_R \phi} , 
\nonumber \\
 w^- &= \sqrt{\frac{r - r_+}{r - r_-}} \ 
\ee^{2 \pi T_L \phi - \frac{t}{2 M}} , \\
 y &= \sqrt{\frac{r_+ - r_-}{r - r_-}} \ 
 \ee^{\pi (T_R + T_L) \phi - \frac{t}{4 M}} , 
\nonumber 
\end{align}
where 
\begin{align}
 T_R &= \frac{r_+-r_-}{4\pi a} , & 
 T_L &= \frac{r_+ + r_-}{4\pi a} . 
\end{align}
Then one can define 
$SL(2,\mathbb R)_R\times SL(2,\mathbb R)_L$ vector fields 
\begin{align}
 H_1 &= i \partial_+ , 
\nonumber 
\\
 H_0 &= i \Big(w^+\partial_+ + \frac{1}{2}y\partial_y\Big) , 
\\
 H_{-1} &= i \Big((w^+)^2\partial_+ 
 + w^+ y \partial_y - y^2 \partial_-\Big) , 
\nonumber
\end{align} 
and 
\begin{align}
 \bar H_1 &= i \partial_- , 
\nonumber 
\\
 \bar H_0 &= i \Big(w^-\partial_- + \frac{1}{2}y\partial_y\Big) , \\
 \bar H_{-1} &= i \Big((w^-)^2\partial_- 
 + w^- y \partial_y - y^2 \partial_+\Big) . 
\nonumber 
\end{align}
In terms of the original coordinates $(t,r,\phi)$, 
these vectors are expressed as 
\begin{align}
 H_1 &= i \ee^{-2\pi T_R \phi} 
 \left(
 \sqrt{\Delta}\partial_r 
 + \frac{1}{2\pi T_R}\frac{r-M}{\sqrt{\Delta}} \partial_\phi
 + \frac{2T_L}{T_R}\frac{Mr-a^2}{\sqrt{\Delta}} \partial_t
 \right) , 
\nonumber 
\\
 H_0 &= i \left(\frac{1}{2\pi T_R}\partial_\phi 
 + 2M\frac{T_L}{T_R}\partial_t\right), \\
 H_{-1} &= i \ee^{2\pi T_R \phi} 
 \left(
 - \sqrt{\Delta}\partial_r 
 + \frac{1}{2\pi T_R}\frac{r-M}{\sqrt{\Delta}} \partial_\phi
 + \frac{2T_L}{T_R}\frac{Mr-a^2}{\sqrt{\Delta}} \partial_t
 \right) , 
\nonumber 
\end{align}
and 
\begin{align}
\bar H_1 &= i \ee^{-2\pi T_L \phi + \frac{t}{2M}} 
 \left(
 \sqrt{\Delta}\partial_r 
 - \frac{a}{\sqrt{\Delta}} \partial_\phi
 - 2M \frac{r}{\sqrt{\Delta}} \partial_t
 \right) , 
\nonumber \\
\bar H_0 &= - 2iM\partial_t, 
\\
\bar H_{-1} &= i \ee^{ 2\pi T_L \phi - \frac{t}{2M}} 
 \left(
 - \sqrt{\Delta}\partial_r 
 - \frac{a}{\sqrt{\Delta}} \partial_\phi
 - 2M \frac{r}{\sqrt{\Delta}} \partial_t
 \right) . 
\nonumber  
\end{align}
The quadratic Casimir 
\begin{equation}
 H^2 = - H_0^2 + \frac{1}{2}(H_1 H_{-1} + H_{-1} H_1), 
\qquad 
 \bar{H}^2 = - \bar{H}_0^2 + \frac{1}{2}(\bar{H}_1 \bar{H}_{-1} 
+ \bar{H}_{-1} \bar{H}_1),  
\label{casimir}
\end{equation}
provide the equation \eqref{RadialEq} as 
\begin{equation}
 H^2\Phi = \bar H^2 \Phi = K\Phi . 
\end{equation}
Therefore, the solution of the scalar field in the Kerr geometry 
in the near region 
forms representations of $SL(2,\mathbb R)$.

\section{Hidden conformal symmetry in the near horizon limit}\label{sec:NH}

We consider the near horizon geometry of the Kerr metric 
which is obtained by defining new coordinates 
\begin{align}
 t &= 2 \epsilon^{-1} a \hat t , &
 r &= a\left(1 + \epsilon \hat r \right) , &
 \phi &= \hat \phi + \frac{t}{2a},  \label{Rescale}
\end{align}
and taking the limit of $\epsilon\to 0$.  

For the extremal case $a=M$, the near horizon geometry becomes  
\begin{equation}
 \dd s^2 = - f_0(\theta) \hat r^2 \dd\hat t^2 
  + f_0(\theta) \frac{\dd\hat r^2}{\hat r^2} 
  + f_\phi(\theta)\left(\dd\hat \phi +  \hat r \dd\hat t \right)^2 
  + f_\theta(\theta)\dd\theta^2 , 
\label{NearHorizon}
\end{equation}
with
\begin{align}
 f_0(\theta) &= f_\theta(\theta) = a^2 \left(1+\cos^2\theta\right) , &
 f_\phi(\theta) &= \frac{4a^2\sin^2\theta}{1+\cos^2\theta} . & 
\label{DetailOfGeometry}
\end{align}
The near horizon geometry has $SL(2,\mathbb R)_R\times U(1)_L$ 
isometries generated by the following four Killing vectors:  
\begin{subequations}
\begin{align}
 \xi_{-1} &= \partial_{\hat t} , & 
 \xi_0 &= \hat{t} \partial_{\hat t} - \hat{r} \partial_{\hat r} , & 
 \xi_{1} &= \left(\hat{t}^2+\frac{1}{\hat{r}^2}\right) \partial_{\hat{t}} 
 - 2 \hat{t}\hat{r} \partial_{\hat r} 
- \frac{2}{\hat{r}}\partial_{\hat\phi} , \label{OriginalSL(2,R)}\\
 \xi_\phi &= \partial_{\hat\phi} , 
\label{OriginalU(1)}
\end{align}
\end{subequations}
where $\xi_{-1}$, $\xi_0$ and $\xi_1$ form the $SL(2,\mathbb R)_R$, 
while $\xi_\phi$ represents the $U(1)_L$ rotational symmetry. 

If non-extremality is infinitesimally small, 
we could take the near horizon limit 
considering the following parameterization 
\begin{equation}
 M = a\left(1 + \epsilon^2\frac{r_0^2}{2}\right), 
\end{equation}
where the parameter $r_0$ measures the deviation from the 
extremality. 
As a result, 
the near horizon geometry becomes 
\begin{equation}
 \dd s^2 = - f_0(\theta) \left(\hat{r}^2-r_0^2\right) \dd \hat{t}^2 
  + f_0(\theta) \frac{\dd\hat{r}^2}{\hat{r}^2-r_0^2} 
  + f_\phi(\theta)\left(\dd\hat{\phi} + \hat{r} \dd\hat{t} \right)^2 
  + f_\theta(\theta)\dd\theta^2 , \label{FiniteTGeometry}
\end{equation}
where $f_0(\theta)$, $f_\phi(\theta)$ and $f_\theta(\theta)$ 
are the same as in \eqref{DetailOfGeometry}. 
The temperature is then given by 
\begin{equation}
 T_H = \frac{r_0}{2\pi} .  
\end{equation}
It is known that this near extremal geometry is related to 
the extremal case via coordinate transformations, and hence 
this geometry inherits the $SL(2,\mathbb R)_R\times U(1)_L$ isometry. 
The Killing vectors of the $SL(2,\mathbb R)_R$ part is given by 
\begin{align}
\label{sl2r}
 \xi_{-1} &= i \ee^{- r_0 \hat{t}} 
\left( \sqrt{\hat{r}^2 - r_0^2} \partial_{\hat r} - \frac{r_0
   \partial_{\hat{\phi}}}{\sqrt{\hat{r}^2 - r_0^2}} 
+ \frac{\hat{r} \partial_{\hat t}}{r_0 \sqrt{\hat{r}^2 -
   r_0^2}} \right), 
\nonumber 
\\
 \xi_0 &= \frac{i}{r_0} \partial_{\hat t} , \\
 \xi_1 &= i \ee^{r_0\hat{t}} 
\left( -\sqrt{\hat{r}^2 - r_0^2} \partial_{\hat r} - \frac{r_0
   \partial_{\hat{\phi}}}{\sqrt{\hat{r}^2 - r_0^2}} 
+ \frac{\hat{r} \partial_{\hat t}}{r_0 \sqrt{\hat{r}^2 -
   r_0^2}} \right) . 
\nonumber  
\end{align}

Let us consider the hidden conformal symmetry 
which appeared in the equation of motion for the 
scalar field.  
In this limit, 
the Laplacian now becomes 
\begin{equation}
 \partial_{\hat r} (\hat{r}^2 - r_{0}^2) \partial_{\hat r} 
- \frac{(\partial_{\hat t} - r_0
   \partial_{\hat\phi})^2}{2 r_0 (\hat{r} - r_0)} 
+ \frac{(\partial_{\hat t} + r_0
   \partial_{\hat\phi})^2}{2 r_0 (\hat{r} + r_0)} 
+ \left( \frac{1}{\sin^2 \theta} - 2 +
   \frac{\sin^2 \theta}{4} \right) \partial_{\hat\phi}^2 
+ \frac{1}{\sin \theta}
   \partial_{\theta} \sin \theta \partial_{\theta} .
\end{equation}
The equation of motion can be separated into 
the radial part and the angular part with a separation
constant $K$:
\begin{subequations}
\begin{equation}
\label{EOMR}
 \left[\partial_{\hat r} (\hat{r}^2 - r_{0}^2) \partial_{\hat r} 
  + \frac{(\omega + r_0 m)^2}{2 r_0 (\hat{r} - r_0)} 
  - \frac{(\omega - r_0 m)^2}{2 r_0 (\hat{r} + r_0)} \right] R(\hat{r})
 = K R(\hat{r}) , 
\end{equation}
and 
\begin{equation}
\label{EOMS}
 \left[
  \frac{1}{\sin \theta}
   \partial_{\theta} \sin \theta \partial_{\theta}  
   - m^2 \left( \frac{1}{\sin^2 \theta} - 2 +
   \frac{\sin^2 \theta}{4} \right)
 \right] S(\theta) = -K S(\theta) ,    
\end{equation}
\end{subequations}
where we have factorized the scalar field 
$\Phi(\hat{t}, \hat{r}, \hat{\phi}, \theta)$ as 
\begin{equation}
 \Phi(\hat{t}, \hat{r}, \hat{\phi}, \theta) 
= \ee^{-i\omega \hat{t} +im\hat{\phi}}R(\hat{r}) S(\theta) . 
\end{equation}
The ``conformal'' coordinates 
are well-defined even in the near horizon limit, 
and we can use it straightforwardly: 
 \begin{align}
\label{AdScoord}
 w^+ &= \sqrt{\frac{r - r_+}{r - r_-}} \ \ee^{2 \pi T_R \phi} 
\longrightarrow
   \sqrt{\frac{\hat{r} - r_0}{\hat{r} + r_0}} \ \ee^{r_0 \hat{t}} , 
\nonumber 
\\
 w^- &= \sqrt{\frac{r - r_+}{r - r_-}} 
\ \ee^{2 \pi T_L \phi - \frac{t}{2 M}} \longrightarrow
   \sqrt{\frac{\hat{r} - r_0}{\hat{r} + r_0}} \ \ee^{\hat{\phi}} ,\\
 y &= \sqrt{\frac{r_+ - r_-}{r - r_-}} \ \ee^{\pi (T_R + T_L) \phi - \frac{t}{4
   M}} \longrightarrow 
\sqrt{\frac{2 r_0}{\hat{r} + r_0}} \ \ee^{\frac{1}{2} (r_0 \hat{t} +
   \hat{\phi})}. 
\nonumber 
 \end{align}
In this limit, the $SL(2,\mathbb R)_R\times SL(2,\mathbb R)_L$ vectors 
can be expressed as 
\begin{subequations}
\begin{align}
 H_1 &= i \ee^{- r_0 \hat{t}} 
\left( \sqrt{\hat{r}^2 - r_0^2} \partial_{\hat r} - \frac{r_0
   \partial_{\hat{\phi}}}{\sqrt{\hat{r}^2 - r_0^2}} 
+ \frac{\hat{r} \partial_{\hat t}}{r_0 \sqrt{\hat{r}^2 -
   r_0^2}} \right) , 
\nonumber 
\\
 H_0 &= \frac{i}{r_0} \partial_{\hat t} , \\
 H_{-1} &= i \ee^{r_0 \hat{t}} 
\left( -\sqrt{\hat{r}^2 - r_0^2} \partial_{\hat r} - \frac{r_0
   \partial_{\hat{\phi}}}{\sqrt{\hat{r}^2 - r_0^2}} 
+ \frac{\hat{r} \partial_{\hat t}}{r_0 \sqrt{\hat{r}^2 -
   r_0^2}} \right) , 
\nonumber 
\end{align}
and 
\begin{align}
 \bar{H}_1 &= i \ee^{- \hat{\phi}} 
\left( \sqrt{\hat{r}^2 - r_0^2} \partial_{\hat r} + \frac{\hat{r}
   \partial_{\hat{\phi}}}{\sqrt{\hat{r}^2 - r_0^2}} 
- \frac{\partial_{\hat t}}{\sqrt{\hat{r}^2 -
   r_0^2}} \right) , 
\nonumber 
\\
 \bar{H}_0 &= i \partial_{\hat{\phi}} , \\
 \bar{H}_{- 1} &= i \ee^{\hat{\phi}} 
\left( - \sqrt{\hat{r}^2 - r_0^2} \partial_{\hat r} + \frac{\hat{r}
   \partial_{\hat{\phi}}}{\sqrt{\hat{r}^2 - r_0^2}} 
- \frac{\partial_{\hat t}}{\sqrt{\hat{r}^2 -
   r_0^2}} \right) . 
\nonumber   
\end{align}
\end{subequations}
One can see the right movers have the same form of (\ref{sl2r}).  
The quadratic Casimir (\ref{casimir})  
gives the equation of motion \eqref{EOMR} as 
\begin{equation}
 H^2\Phi = \bar H^2 \Phi = K\Phi . 
\end{equation}
Generally $K$ possibly depends on 
the angular momentum along $\phi$-direction. 
In the near horizon limit, the conditions 
for the hidden conformal symmetry can be interpreted as follows. 
The original conditions of $\omega\ll 1/M$ and $\omega\ll 1/r$ 
can be roughly understood as small $\omega$ condition. 
In order to take $\omega$ to be small in the original geometry, 
the angular momentum $m$ should be small in the near horizon geometry. 
For $m\ll l$, the potential term in \eqref{EOMS} 
can be approximated as
\begin{equation}
 m^2 \left( \frac{1}{\sin^2 \theta} - 2 + \frac{\sin^2 \theta}{4} \right) 
  \sim \frac{m^2}{\sin^2\theta} . 
\end{equation}
Then, \eqref{EOMS} reduces to that for 2-sphere, 
and $K$ becomes 
\begin{equation}
 K \to l(l+1)
\end{equation}
in this limit. 

We can consider the hidden conformal symmetry 
for the near horizon extremal Kerr geometry 
by taking the $r_0\to 0$ limit. 
In this limit, the ``conformal'' coordinates can be expressed as 
 \begin{align}
\label{ConfCoordExt}
 w^+ &= \hat{t} - \frac{1}{\hat{r}} , 
\nonumber 
\\
 w^- &= \ee^{\hat{\phi}} , \\
 y &= \sqrt{\frac{2}{\hat{r}}} \ \ee^{\frac{1}{2} \hat{\phi}} ,
\nonumber  
 \end{align}
after an appropriate rescaling. 
Then, $SL(2,\mathbb R)_R\times SL(2,\mathbb R)_L$ vectors become
\begin{subequations}\label{SL(2,R)}
\begin{align}
 H_1 &= i\partial_+ = i\partial_{\hat t} , 
\nonumber 
\\
 H_0 &= i\Big(w^+\partial_+ + \frac{1}{2} y \partial_y\Big) 
 = i\left(\hat{t} \partial_{\hat t} - \hat{r} \partial_{\hat r}\right) , \\
 H_{- 1} &= i\Big((w^+)^2 \partial_+ + w^+ y \partial_y - y^2 \partial_-\Big)
 = i\left(\Big(  \hat{t}^2 + \frac{1}{\hat{r}^2} \Big) \partial_{\hat t}
  - 2 \hat{t}\hat{r} \partial_{\hat r} - \frac{2}{\hat{r}} 
\partial_{\hat{\phi}}\right) , 
\nonumber 
\end{align}
and
\begin{align}
  \bar H_1 &= i\partial_- = i\ee^{- \hat{\phi}} 
\left( \hat{r} \partial_{\hat r} +
   \partial_{\hat{\phi}} - \frac{1}{\hat{r}} \partial_{\hat t} 
\right) , 
\nonumber 
\\
 \bar{H}_0 &= i\Big(w^- \partial_- + \frac{1}{2} y \partial_y\Big)
 = i\partial_{\hat{\phi}} , \\
 \bar{H}_{- 1} &= i\Big(
(w^-)^2 \partial_- + w^- y \partial_y - y^2 \partial_+\Big) =
   i\ee^{\hat{\phi}} 
\left( - \hat{r} \partial_{\hat r} 
+ \partial_{\hat{\phi}} - \frac{1}{\hat{r}} \partial_{\hat t}
   \right) .
\nonumber  
\end{align}
\end{subequations}
The right movers reproduce (\ref{OriginalSL(2,R)}). 
The quadratic Casimir  
\begin{equation}
 H^2 = \bar{H}^2 = \partial_{\hat{r}} {\hat r}^2 \partial_{\hat r} 
- \frac{(\partial_{\hat t} - {\hat r}
 \partial_{\hat{\phi}})^2}{\hat{r}^2} + \partial_{\hat{\phi}}^2 
\end{equation}
gives the equation of motion for $R(\hat{r})$ as in the near extremal case. 

Before closing the section, 
we consider how the hidden symmetry can be understood 
in terms of the geometry. 
The eigenvalue of \eqref{EOMS} contains a contribution 
from the angular momentum $m$. 
Even in the case of $m \ll l$, 
$K$ has contributions from $m$.  
However, if we consider the point $\theta_0$ 
which satisfies the following condition for the potential terms 
in~\eqref{EOMS}, 
\begin{equation}
 \left( \frac{1}{\sin^2 \theta_0} 
- 2 + \frac{\sin^2 \theta_0}{4} \right) = 0 , 
\end{equation}
$m$ does not contribute to the equation 
and the $\phi$-direction completely decouples from the $\theta$-direction. 
Near the point $\theta_0$, the geometry can be approximated as
\begin{equation}
 \dd s^2 = f_0 (\theta_0) 
  \left[ - (\hat{r}^2 - r_0^2) \dd {\hat t}^2 
+ \frac{\dd \hat{r}^2}{\hat{r}^2 -
   r_0^2} + (\dd \hat{\phi} + \hat{r} \dd \hat{t})^2 + \dd\theta^2 \right]. 
\end{equation}
Using the ``conformal'' coordinates this can be rewritten as
\begin{equation}
  \dd s^2 = 4f_0(\theta_0)\left[
  \frac{\dd w^+ \dd w^-}{y^2} + \frac{\dd y^2}{y^2} 
+\frac{\dd\theta^2}{4}\right]. 
\end{equation}
The hidden conformal symmetry becomes 
an exact symmetry on the slice $\theta=\theta_0$. 

In the following sections, we work with the near horizon coordinates 
\eqref{ConfCoordExt} and omit the hat `` $\hat\ $ '' on the 
coordinates.

\section{Virasoro algebra}\label{sec:Virasoro}

We have seen that the Laplacian can be expressed by 
the quadratic Casimir of the $SL(2,\mathbb R)$ vectors. 
Here, we assume that these $SL(2,\mathbb R)$ vectors 
can be extended to the Virasoro generators 
and have the same forms to those in AdS$_3$. 
By using the ``conformal'' coordinates, 
the Virasoro generators might be 
\begin{subequations}\label{VirasoroExp}
 \begin{align}
 \xi_n &= i \left( (w^+)^{n + 1} \partial_+ + \frac{n + 1}{2} (w^+)^n y
   \partial_y - \frac{n (n + 1)}{2} (w^+)^{n - 1} y^2 \partial_- \right) , \\
 \bar{\xi}_n &= i \left( (w^-)^{n + 1} \partial_- + \frac{n + 1}{2} (w^-)^n y
   \partial_y - \frac{n (n + 1)}{2} (w^-)^{n - 1} y^2 \partial_+ \right) . 
 \end{align}
\end{subequations}
These vectors can be expressed 
by using arbitrary functions $f_+(w^+)$ and $f_-(w^-)$ as 
\begin{subequations}
\label{VirasoroConf}
 \begin{align}
 \xi &= f_+(w^+) \partial_+ + \frac{1}{2} f_+'(w^+) y \partial_y
 - \frac{1}{2}  f_+'' (w^+) y^2 \partial_- , \\
 \bar{\xi} &= f_-(w^-) \partial_- + \frac{1}{2} f_-'(w^-) y \partial_y -
   \frac{1}{2} f_-''(w^-) y^2 \partial_+ . 
 \end{align}
\end{subequations}
The vectors \eqref{VirasoroExp} are given as 
the Taylor expansion of these vectors. 
By using the near horizon coordinates, 
the vectors \eqref{VirasoroConf} are written as 
\begin{subequations}
\label{VectorExtremal}
 \begin{align}
 \xi &= \left[ f_+ (t - \frac{1}{r}) + \frac{1}{r} f_+' (t - \frac{1}{r}) +
   \frac{1}{r^2} f_+'' (t - \frac{1}{r}) \right] \partial_t 
 \notag\\&\quad - \left[ r f_+' (t -
   \frac{1}{r}) + f_+'' (t - \frac{1}{r}) \right] \partial_r - \frac{1}{r} f_+''
   (t - \frac{1}{r}) \partial_{\phi} 
\nonumber 
\\
 &= 
  \Big(
   f_R(t) + \frac{f_R''(t)}{2r^2} +\mathcal O(r^{-3}) 
  \Big)
  \partial_t 
  + 
  \Big(
   - r f_R'(t) + \frac{f_R'''(t)}{2r} + \mathcal O(r^{-2}) 
  \Big)
  \partial_r 
 \notag\\&\quad
  +\Big( - \frac{f_R''(t)}{r}
  + \mathcal O(r^{-2}) 
  \Big)
  \partial_\phi , 
\label{RightVirasoro}\\
 \bar{\xi} &= f_L(\phi) \partial_{\phi} - r f_L'(\phi) \partial_r
 - \frac{1}{r}  f_L''(\phi) \partial_t , 
\label{LeftVirasoro}
 \end{align}
\end{subequations}
where $f_R(t):=f_+(t)$ and $f_L(\phi):=\ee^{-\phi}f_-(\ee^\phi) $. 
The vectors for right movers are equivalent 
to the asymptotic Killing vectors in \cite{mty} at least up to 
the given orders in (\ref{RightVirasoro}). 
However, the vectors for left movers have an additional term in
$t$-direction at $\mathcal O(r^{-1})$. 
This term was excluded in the original 
asymptotic Killing vector which was derived in \cite{ghss}. 

In the following sections, we use the vector fields 
(\ref{VectorExtremal}). 
For the precise calculations, 
we define the right and left movers $\xi_n$ and $\bar{\xi}_n$ 
through the expressions (\ref{VectorExtremal}) 
with $f_R(t)=it^{n+1}$ and $f_L(\phi)=\ee^{in\phi}$. 
The subsets $(\xi_{-1}, \xi_{0}, \xi_{1})$ and 
$(\bar{\xi}_{-1}, \bar{\xi}_{0}, \bar{\xi}_{1})$ form 
$SL(2, \mathbb R)_R$ and $SL(2, \mathbb R)_L$, respectively, 
but \eqref{SL(2,R)} are given by 
$(\xi_{-1}, \xi_{0}, \xi_{1})$ and 
$(i\bar{\xi}_{-i}, i\bar{\xi}_{0}, i\bar{\xi}_{i})$.

\section{Central charge from the asymptotic charge}
\label{sec:Central}

In order to calculate the central charge, 
we first consider the asymptotic charge 
which is given by the covariant formulation~\cite{bb,bc}. 
The asymptotic charge is defined 
as the deviation of the charge from the 
background $\bar{g}_{\mu\nu}$. 
For infinitesimal perturbations $h_{\mu\nu}$, 
that is given by 
\begin{equation}
 Q^{\rm A}_\xi[h] 
  = \frac{1}{8\pi G_N}
  \!\int_{\partial\Sigma_\infty}\!\!\!\!\! 
  k_{\xi}[h,\bar g] , 
  \label{asymcharge}
\end{equation}
where the integration is taken over the boundary of a time slice.  
The two-form $k_\xi$ is defined by 
\begin{equation}
 k_{\xi}[h,\bar g] 
 = \frac{\sqrt{-\bar g}}{4}\epsilon_{\mu\nu\rho\sigma}
 \tilde k_{\xi}^{\mu\nu}[h,\bar g]\, \dd x^\rho \wedge \dd x^\sigma , 
\end{equation}
with
\begin{align}
 \tilde k_{\xi}^{\mu\nu}[h,\bar g]
 = \frac{1}{2}\Bigl[
& 
 \xi^\mu D^\nu h 
 - \xi^\mu D_\lambda h^{\lambda\nu} 
 + \left(D^\mu h^{\nu\lambda}\right)\xi_\lambda 
 + \frac{1}{2} h D^\mu \xi^\nu 
 \notag\\
&
 - h^{\mu\lambda}D_\lambda\xi^\nu 
 + \frac{1}{2}h^{\mu\lambda}
 \left(D^\nu\xi_\lambda+D_\lambda\xi^\nu\right) 
 - (\mu\leftrightarrow\nu) \Bigr] , 
\end{align}
where $D_\mu$ is a covariant derivative on the background geometry, 
and we denote $\bar g = \det \bar g_{\mu\nu}$ 
and $h = \bar g^{\mu\nu}h_{\mu\nu}$. 
The central charge 
can be calculated by varying the charge
\begin{equation}
\delta_{\zeta} Q^{\rm A}_{\xi} 
=
 \frac{1}{8\pi G_N}\!\int_{\partial\Sigma_\infty}\!\!\!\!\!
 k_{\xi}[\pounds_{\zeta}\bar g, \bar g]
 + \frac{1}{8\pi G_N}\!\int_{\partial\Sigma_\infty}\!\!\!\!\!
 k_{\xi}[\pounds_{\zeta} h , \bar g] ,  
\end{equation}
where $\pounds_{\zeta}$ stands for the Lie derivative along the vector 
field $\zeta$. 
The second term which is proportional to the perturbation $h$ 
gives the usual transformation. 
The first term is an additional constant term, 
which provides the central charge of the algebra. 

Let us estimate the central charge for the vectors 
(\ref{VectorExtremal}).  
For the right movers, this term vanishes: 
\begin{equation}
 \frac{1}{8\pi G_N}\!\int_{\partial\Sigma_\infty}\!\!\!\!\!
 k_{\xi_n}[\pounds_{\xi_m}\bar g, \bar g] = 0 .  
\label{cr}
\end{equation}
This implies $c_R=0$ for the right movers of Virasoro algebra.
For the left movers, we obtain  
\begin{equation}
 \frac{1}{8\pi G_N}\!\int_{\partial\Sigma_\infty}\!\!\!\!\!
 k_{\bar \xi_n}[\pounds_{\bar \xi_m}\bar g, \bar g]
 =  -2i\frac{a^2}{G_N}\delta_{n+m,0}\big(n+n^3-\Lambda n^5\big) . 
\label{cl}
\end{equation}
The term ${\cal O}(n)$ could be absorbed by taking an appropriate 
definition of $\bar{Q}^{\rm A}_0$, while the term ${\cal O}(n^3)$ 
contributes to the central charge. 
In addition to these, we here get an unusual term 
in $\mathcal O(n^5)$ whose coefficient $\Lambda$ is an 
infinitely large constant. 
Neglecting this term, one may read off the central charge 
$c_L = \frac{24 a^2}{G_N}= 24J$.  
However this is twice of the original study~\cite{ghss}. 
In order to understand the origin of the extra term of $\mathcal O(n^5)$, 
we shall consider the quasi-local charge \'a la asymptotic charge.  

The quasi-local charge is defined by using 
the energy-momentum tensor \cite{by} (see also \cite{bk}), 
which is defined in the same form to the GKPW relation~\cite{gkp,w}:
\begin{equation}
 T^{\mu\nu}
  = \frac{2}{\sqrt{-\gamma}}
  \frac{\delta S_{\text{grav}}}{\delta\gamma_{\mu\nu}} ,  
\end{equation}
where $\gamma_{\mu\nu}$ is the metric on the boundary. 
The gravitational action $S_{\text{grav}}$ 
is the Einstein-Hilbert action 
with the Gibbons-Hawking term, 
\begin{equation}
 S_{\text{grav}} 
  = \frac{1}{16\pi G_N} \!\int_{\mathcal M}\!\! \dd^4 x\sqrt{-g} R 
  + \frac{1}{8\pi G_N}\!\int_{\partial\mathcal M}\!\!\!\! 
  \dd^3 x \sqrt{-\gamma} K, 
\end{equation}
where $K=K^\mu_\mu$ and $K_{\mu\nu}$ is the extrinsic curvature. 
Let us consider 
following decomposition of the metric: 
\begin{equation}
 \dd s^2 = N^2 \dd r^2 
+ \gamma_{\mu\nu}(\dd x^\mu + N^\mu \dd r)(\dd x^\nu + N^\nu \dd r). 
\end{equation}
We focus on the variation of the action with respect to 
the induced metric $\gamma_{\mu\nu}$. 
Since we only consider an on-shell action, 
the variation of the action is reduced to boundary terms, 
\begin{equation}
 \delta S = \!\int_{\partial\mathcal M}\!\!\!\! \dd^3 x\ 
  \pi^{\mu\nu} \delta\gamma_{\mu\nu} , 
\end{equation}
where $\pi^{\mu\nu}$ is the conjugate momentum of $\gamma_{\mu\nu}$. 
Then the energy-momentum tensor 
on the boundary is expressed as 
\begin{equation}
 T^{\mu\nu} 
= \frac{2}{\sqrt{-\gamma}}\,\pi^{\mu\nu} 
= \frac{1}{8\pi G_N} 
  \left(K^{\mu\nu}-\gamma^{\mu\nu}K\right) . 
\end{equation}
We next consider the ADM decomposition of the boundary 
metric $\gamma_{\mu\nu}$, 
\begin{equation}
 \dd s^2_{\partial\mathcal M} 
  = -N^2_{\partial\Sigma} \dd t^2 
  + \sigma_{ab} (\dd x^a + N^a_{\partial\Sigma} \dd t)
  (\dd x^b + N^b_{\partial\Sigma} \dd t) ,
\end{equation}
where $\Sigma$ is a time slice and 
$\partial\Sigma$ is that on the boundary. 
Then the quasi-local charge on the boundary can be defined by 
\begin{equation}
 Q^{\rm QL}_\xi = \!\int_{\partial\Sigma}\!\! 
\dd^2 x\sqrt{\sigma}\,
  u^\mu T_{\mu\nu} \xi^\nu , 
\label{QLCharge}
\end{equation}
where $u^\mu$ is a future pointing timelike 
unit normal to $\partial\Sigma$. 
In order to define the charge, 
we need to fix the reference, 
and take the difference of the charge form the reference. 
We choose it in the following way, such that 
the quasi-local charge reproduces 
the asymptotic charge. 
We take the background geometry as the reference geometry, 
since the asymptotic charge is defined as 
its deviation from the background. 
Here, we define the charge by taking the 
difference of the charge itself. 

The central charge can be calculated from 
the variation of the charge: 
\begin{equation}
 \delta_\zeta Q_\xi^{\rm QL}[\bar g + h] 
  = Q_\xi^{\rm QL}[\bar g + \pounds_\zeta \bar g] 
    - Q_\xi^{\rm QL}[\bar g]
    + (\text{terms which depends on }h). \label{VariationQ}
\end{equation}
Then, we can read off the central charge from the first two terms 
in the r.h.s..  
Using this definition, we obtain 
\begin{equation}
 Q_{\xi_n}^{\rm QL}[\bar g + \pounds_{\xi_m} \bar g] 
    - Q_{\xi_n}^{\rm QL}[\bar g] = 0 ,  
\end{equation}
for the right movers and 
\begin{equation}
 Q_{\bar\xi_n}^{\rm QL}[\bar g + \pounds_{\bar\xi_m} \bar g] 
    - Q_{\bar\xi_n}^{\rm QL}[\bar g]
    = -2i\frac{a^2}{G_N}\delta_{n+m,0}\big(n+n^3-\Lambda n^5\big) , 
\end{equation}
for the left movers. 
As we promised, these results reproduce those of the asymptotic 
charge (\ref{cr}) and (\ref{cl}). 
We can see the origin of this divergent term 
by looking at the definition of the quasi-local charge. 
The definition of the quasi-local charge is 
almost equivalent to the charge defined by using the GKPW relation. 
The quasi-local energy-momentum tensor $T_{\mu\nu}$ is 
equivalent to the expectation value of the energy-momentum tensor 
$\langle T_{\mu\nu} \rangle_{\rm CFT}$ in the GKPW relation. 
However, the Killing vector $\xi$ is different 
between the quasi-local charge and the charge in the GKPW relation. 
In the above calculation, we simply used the vectors 
\eqref{VectorExtremal}. 
However, the Killing vectors are 
$\xi = f(z)\partial$ and $\bar\xi = \bar f(\bar z)\bar\partial$ 
for ordinary CFT case, and hence we should use these vectors 
for the GKPW relation. 
Now, the origin of the order $\mathcal O(n^5)$ term is clear. 
The anomaly gives $\mathcal O(n^3)$ term 
to transform of the energy-momentum tensor, and 
the extra term in $\bar\xi$, $f_L''(\phi)/r^2 \partial_t$ is of order $n^2$. 
Then, we get the $\mathcal O(n^5)$ term in the central charge. 

In order to obtain the appropriate algebraic relation 
without the unexpected terms,  
we should use only the leading term of the vectors \eqref{VectorExtremal}, 
i.e.\ vectors on the boundary,  
$\xi_{\mathrm B n} = t^{n+1}\partial_t$ for 
the right movers and 
$\bar\xi_{\mathrm Bn} = \ee^{in\phi}\partial_\phi$ for the 
left movers. 
By using these definitions, we obtain
\begin{equation}
 Q_{\xi_{\mathrm B n}}^{\rm QL}[\bar g + \pounds_{\xi_m} \bar g] 
    - Q_{\xi_{\mathrm B n}}^{\rm QL}[\bar g] = 0 ,
\end{equation}
for the right movers and 
\begin{equation}
 Q_{\bar\xi_{\mathrm B n}}^{\rm QL}[\bar g + \pounds_{\bar\xi_m} \bar g] 
    - Q_{\bar\xi_{\mathrm B n}}^{\rm QL}[\bar g]
    = -2i\frac{a^2}{G_N}\delta_{n+m,0}\big(n+n^3\big) ,  
\end{equation}
for the left movers. 
These results ensure appropriate algebraic relations, 
and the central charges would be $c_R = 0$ and $c_L = 24J$. 

It should be noted that 
the original asymptotic Killing vector for left movers 
\begin{equation}
 \xi^{(0)} = f(\phi)\partial_\phi - r f'(\phi) 
\partial_r \label{VectorOriginal}
\end{equation}
gives $c_L = 12 J$ 
even if we use the definition of the quasi-local charge: 
\begin{equation}
 Q_{\xi_{n}^{(0)}}^{\rm QL}[\bar g + \pounds_{\xi_m^{(0)}} \bar g] 
    - Q_{\xi_{n}^{(0)}}^{\rm QL}[\bar g]
    = -i\frac{a^2}{G_N}\delta_{n+m,0}\big(n+n^3\big) . 
\end{equation}

\section{Relation to the entropy}\label{sec:Entropy}

In the previous section, we calculated 
the central charge. 
However, the result is different from that in \cite{ghss}, 
and hence it does not reproduce 
the Bekenstein-Hawking entropy. 
In this section, we consider a different relation 
of Kerr/CFT correspondence following our previous work~\cite{mty}. 

As we saw in the previous section, 
in order to define the quasi-local charges, 
there is an ambiguity how to take 
the reference and the deviation from it. 
We here use the following definition of the quasi-local charge~\cite{mty}:  
\begin{equation}
 {\cal Q}^{\rm QL}_\xi = \!\int_{\partial\Sigma}\!\! \dd^2 x\sqrt{\sigma}\,
  u^\mu \tau_{\mu\nu} \xi_\mathrm{B}^\nu, \label{redefQLCharge}
\end{equation}
where we define the quasi-local energy-momentum tensor 
by its deviation from the background, 
\begin{equation}
 \tau^{\mu\nu}[h] = 
  T^{\mu\nu}\Bigr|_{g=\bar g + h} - T^{\mu\nu}\Bigr|_{g=\bar g} , 
\end{equation}
and we use the leading term of the vectors (\ref{VectorExtremal}) 
i.e.\ $\xi_{B}$ and $\bar{\xi}_{B}$ to close the algebra. 
The central charge can be derived from 
the anomaly of the energy-momentum tensor.  
In \eqref{VariationQ}, $\delta Q_\xi$ contains 
not only the anomaly of the energy-momentum tensor 
but metric gives additional anomalous terms. 
By using \eqref{redefQLCharge} instead, we can pick up 
the anomaly of the energy-momentum tensor,
\begin{equation}
 \delta T^{\mu\nu} = \tau^{\mu\nu}[\pounds_\xi\bar g] . 
\end{equation}
Then, we can read off the central charge from 
\begin{equation}
 \delta_{\xi_{n}} {\cal Q}_{\xi_{m}}^{\rm QL} 
  = \int_{\partial\Sigma}\!\! \dd^2 x\sqrt{\sigma}\,
  u^\mu \tau_{\mu\nu}[\pounds_{\xi_{n}} \bar g] \xi_{\mathrm{B}m}^\nu . 
\label{redefQLCCharge}
\end{equation}

Let us discuss the entropy through the Cardy formula, 
\begin{equation}
S_R=2\pi\sqrt{\frac{c_R L_0}{6}}, 
\qquad 
S_L=2\pi\sqrt{\frac{c_L \bar{L}_0}{6}}.  
\label{cardy}
\end{equation}
For the right movers, 
the vectors \eqref{RightVirasoro} 
are equivalent to the asymptotic Killing vectors in \cite{mty}. 
The central charge and $L_0$ are calculated as 
\begin{align}
 c_R &= \frac{12\epsilon a^2}{G_N} , &
 L_0 &= \frac{\epsilon a^2 r_0^2}{2G_N} , 
\end{align}
and the entropy is estimated as:  
\begin{equation}
 S_R = \frac{2\pi \epsilon a^2 r_0}{G_N} ,  
\end{equation}
where we introduce a regularization by putting the boundary 
$\partial\Sigma$ at $r = \epsilon^{-1}$. 
For the left movers, 
evaluating (\ref{redefQLCCharge}), 
\begin{align}
 \int_{\partial\Sigma}\!\! \dd^2 x\sqrt{\sigma}\,
  u^\mu \tau_{\mu\nu}[\pounds_{\bar{\xi}_n}\bar g] 
\bar{\xi}_{\mathrm{B}m}^\nu 
 = -i\delta_{n+m,0}\big(n+n^3\big) \frac{\epsilon^2a^2 r_0^2}{G_N} ,  
\end{align}
we can read off the central charge: 
\begin{equation}
 c_L = \frac{12 \epsilon^2 a^2 r_0^2}{G_N} . 
\end{equation}
For the left movers, $\bar L_0$ corresponds 
to the angular momentum $J$, but its deviation form the extremality: 
\begin{equation}
 \bar L_0 = \frac{\epsilon^2 a^2 r_0^2}{2 G_N} . 
\end{equation}
Then, using the Cardy formula (\ref{cardy}), we obtain 
\begin{equation}
 S_L = 
\frac{2\pi \epsilon^2 a^2
  r_0^2}{G_N} . 
\end{equation}
Expanding the Bekenstein-Hawking entropy (\ref{bhe}) as 
\begin{equation}
 S_{\rm BH} = \frac{2\pi a^2}{G_N}
  \Big(1 + \epsilon r_0 + \epsilon^2 r_0^2 + \cdots\Big) , 
\end{equation}
we can observe that the right movers reproduce the second term 
and the left movers reproduce the third term. 
In the case of the asymptotic Killing vector,  
the Cardy formula reproduces the entropy 
at the extremality (see Appendix A).  
In the present case, however, due to the additional 
term of $\mathcal O(1/r)$ in 
the vectors \eqref{LeftVirasoro}, 
the formula does not give the entropy 
at the extremality, 
but reproduce only its deviation from the extremality. 

This is similar to the case of the AdS/CFT correspondence. 
In the AdS$_5$/CFT$_4$ correspondence of the D3-brane, for example, 
we can see the correspondence between the 
energy density in the CFT and the mass in the AdS space. 
Although the mass in the AdS is the ADM mass of the black 3-brane, 
we have to take the deviation from the extremality 
to make a connection to the energy in CFT. 
The mass at the extremality is never reproduced by the CFT 
since this is the tension of the D-brane itself and 
the CFT is only the fluctuations on the D-brane. 

If we consider the near horizon geometry of extremal Kerr (NHEK) 
as an analogy of the extremal BTZ black hole, 
the entropy at the extremality should be reproduced 
by the corresponding CFT. 
However, it is rather natural to treat the NHEK geometry 
as a correspondent of the pure AdS space. 
Then, it is not strange that 
physical quantity in Kerr black hole at the extremality 
cannot be reproduced by the corresponding CFT. 
The CFT might come from the fluctuations 
on the Kerr black hole (or its horizon) 
and reproduce only the deviation of the physical quantity 
from the extremality.

\section{Discussions}\label{sec:Disc}

In this paper, we have studied the hidden conformal symmetry 
in the near horizon limit. 
For the near extremal case, we can take the well-defined 
near horizon limit for the ``conformal'' coordinates and 
$SL(2,\mathbb R)_R\times SL(2,\mathbb R)_L$ vectors. 
We do not need to introduce any rescaling for the vectors 
to take the limit. 
By using a simple generalization to the Virasoro algebra, 
we obtained two sets of vectors. 
These vectors agree with the asymptotic Killing vectors 
obtained in \cite{ghss} and \cite{mty} at the leading order. 
For the right movers, the vectors agree with 
the asymptotic Killing vectors even at the higher order. 
However, the vectors for left movers receive a correction 
in the $t$-component at $\mathcal O(1/r)$. 
We calculated the central charge by using 
the asymptotic charge and the quasi-local charge 
with suitable definition.  
This did not lead to the expected values of $c_L = c_R = 12 J$. 
By using the asymptotic charge, 
the central charges became $c_R = 0$ and $c_L = 24J$, 
and using the quasi-local charge, 
we obtain $c_R = c_L = 0$ at the extremality, 
and $c_R = 12 \epsilon a^2/G_N$ and 
$c_L = 12 \epsilon^2 a^2 r_0^2/G_N$ 
as near-extremal corrections. 

This result requires further discussions. 
The first point to be considered is 
the extension to the Virasoro algebra. 
The ``conformal'' coordinates give the 
Laplacian on AdS$_3$ in Poincare coordinates. 
Hence we simply used the asymptotic Killing vectors 
in AdS$_3$ as the enhanced vectors 
in the ``conformal'' coordinates. 
One may introduce a different extension of 
the $SL(2,\mathbb R)$ Killing vectors to the Virasoro algebra. 
However, it should be noted that 
the disagreement for the left central charge 
comes from the $O(1/r)$ correction 
in the $t$-component of the vectors for the Virasoro algebra. 
This correction appears even in the $SL(2,\mathbb R)_L$ vectors, 
for which we introduce no extensions. 
This correction is excluded in the original definition 
of the asymptotic Killing vectors \cite{ghss}. 
In other words, the original asymptotic Killing vectors 
do not have $t$-component, and such vectors 
never reproduce the Laplacian. 

Next, we should discuss the definition of the charge. 
By using the covariant definition of the asymptotic charge, 
we obtained the central extension 
with an unexpected term of $\mathcal O(n^5)$. 
In fact, the vectors of the hidden conformal symmetry are 
not the Killing vectors of the asymptotic symmetry, 
and hence the asymptotic charge is possibly not 
appropriate to define the charge of these vectors. 
In order to obtain well-defined central extension, 
we need to use the quasi-local charge. 
However, there is an ambiguity how to take 
the difference from the charge in the reference geometry.%
\footnote{
For example, if we use the following definition: 
$$
 Q^{\rm QL}_\xi = \!\int_{\partial\Sigma}\!\! 
\dd^2 x\sqrt{\sigma}\, 
 \left[u^\mu T_{\mu\nu} \xi^\nu \bigr|_{g=\bar g+h} 
 - u^\mu T_{\mu\nu} \xi^\nu \bigr|_{g=\bar g} \right] , 
$$
the central charge becomes $c_L = 12 J$ for 
\eqref{VectorExtremal} but $c_L = 6 J$ for \eqref{VectorOriginal}. 
We do not pursue this direction since it is a quite arbitrary definition. 
} 
We considered two definitions.
One is taken such that 
the result reproduce that of the asymptotic charge. 
As an another definition, we took the difference 
of the quasi-local energy-momentum tensor $T^{\mu\nu}$. 
By replacing the Killing vector $\xi$ on the boundary 
by $\xi_B$ which picks up only the leading term, 
the term of $\mathcal O(n^5)$ does not appear. 
Then, the former definition does not reproduce the entropy, 
but the latter gives the non-extremal correction of the entropy. 
The result for the latter definition 
is similar to the AdS/CFT correspondence. 
For example, AdS$_5$ is near horizon geometry 
of the extremal black 3-brane. 
CFT does not reproduce the ADM mass of the extremal black 3-brane 
but gives only its deviation from the extremality. 
Hence, for the hidden conformal symmetry, 
we should treat NHEK as an analogy of AdS itself, 
not as an analogy of the extremal BTZ black hole. 
The counter term method \cite{bk}, which is used in \cite{cl} 
can be an alternative definition of how to take 
the difference form the reference. 
However, $\theta$-dependence of the metric 
makes it difficult to cancel divergent part in 
the energy momentum tensor. 
In \cite{cl}, this can be done by integrating out 
$\phi$ and $\theta$ dependence, 
but this method cannot used to calculate the left central charge 
whose vectors depend on $\phi$. 

We should comment on the right central charge. 
Our result $c_R = 0$ is consistent 
with the result of \cite{mty}, 
but different from that in \cite{cl} and \cite{cms}, i.e.\ $c_R = 12J$. 
This discrepancy comes from the definition of the central charge. 
In \cite{cl}, the central charge is defined 
from the anomalous transformation of the energy-momentum: 
$$
 \delta T_{tt} \sim - \frac{c_{\rm CL}}{12} f'''(t)  +\cdots . 
$$
On the other hand, we have defined the central charge 
as the central extension of the algebra: 
$$
 \delta Q \sim - \frac{c_{\rm ours}}{12}(n+n^3) . 
$$
Since the charge is defined by 
the integration of the energy-momentum tensor \eqref{QLCharge}, 
we can see the following schematic relation for these two central charges: 
$$
 c_{\rm ours} \sim 
  \!\int\! \dd^2 x \sqrt{\sigma} u^t \xi^t c_{\rm CL} 
  = \!\int\! \sqrt{-\gamma} \gamma^{tt} c_{\rm CL} . 
$$
For the near horizon geometry of the Kerr black hole, 
we have 
$$
 \sqrt{-\gamma} \gamma^{tt} \sim u^t \sim \frac{1}{r} . 
$$
Since we are considering the boundary at $r\to\infty$, 
our central charge becomes zero 
even though that in \cite{cl} is finite. 
Of course, if we assume only the correspondence of 
the energy-momentum and take 
the flat metric for the boundary field theory, 
the definition of \cite{cl} 
is related to the generators on the boundary. 
However, this treatment leads to disagreement 
of the charges in gravity side and CFT side, 
or we need to take into account effects of 
the gauge transformation. 

Our central charges can be continuously connected 
to generic non-extremal cases. 
When we take the near horizon limit, we have not introduced 
any rescaling for the vectors. 
The charges are defined as the integration on the 
$(\phi,\theta)$-plane with $t$ fixed. 
This integration is not rescaled in the near horizon limit, 
and fixed $t$ plane is equivalent to fixed $\hat t$ plane. 
It is different that the boundary of the near horizon geometry 
is located inside the near horizon region. 
By using the definition of the quasi-local charge, 
this can be interpreted as follows; 
we take into account only 
contributions inside the near horizon region 
but do not consider those far from the black hole. 
However, we can assume that there are only small 
contributions in the asymptotically flat region, 
and this can be neglected in the near extremal case. 
Then, the central charges for generic non-extremal cases 
are expected to be continuously connected to our result. 
If we could calculate the central charge 
without taking the near horizon limit, 
the result would agree with that in this paper 
for the near extremal. 

In this paper, we have considered the hidden conformal symmetry 
in the near horizon limit. 
However it seems that there is no simple relation to 
the Virasoro algebras of the asymptotic symmetry. 
We could reproduce a part of the entropy by using 
the quasi-local charge, 
but the asymptotic charge does not give the entropy. 
This result implies that the hidden conformal symmetry 
should be treated in a different framework from 
the asymptotic symmetry. 
In this paper, we used the same vectors to the AdS$_3$. 
It would be interesting to consider more explicit 
formulation to extend the hidden conformal symmetry to 
the Virasoro algebra. 
This is left for future studies.

\vspace*{5mm}

\noindent
 {\large{\bf Acknowledgements}}

TT acknowledges the Max Planck Society (MPG),
the Korea Ministry of Education, Science, Technology (MEST), 
Gyeongsangbuk-Do and Pohang City for the support of the 
Independent Junior Research Group at the Asia Pacific Center for 
Theoretical Physics (APCTP). 
CY is supported by JSPS Grant-in-Aid for Creative Scientific 
Research No.~19GS0219.

\vspace*{2mm}

\appendix

\section{Central charge from the quasi-local \\ charge}

In this appendix, we derive the central charge 
for the asymptotic Killing vectors 
by using the quasi-local charge defined in Section \ref{sec:Entropy}, 
and show the relation to the entropy. 
As explained in~\cite{ghss}, 
although the original Cardy formula (\ref{cardy}) 
does not reproduce the entropy, 
we could use the following formula for the entropy 
\begin{equation}
 S_L = \frac{\pi^2}{3} c_L T_L , 
\qquad 
 S_R = \frac{\pi^2}{3} c_R T_R .   
\label{Cardy2}
\end{equation}

For left movers, the asymptotic Killing vector 
is given by \eqref{VectorOriginal} \cite{ghss}. 
Introducing the regularization parameter $\epsilon$ 
in the same manner in the main text, 
we can obtain 
\begin{align}
 \int_{\partial\Sigma}\!\! \dd^2 x\sqrt{\sigma}\,
  u^\mu \tau_{\mu\nu}[\pounds_{\xi_n^{(0)}}\bar g] \xi_m^{(0)\nu} 
 = -i\delta_{n+m,0} \frac{a^2}{G_N}
 \Big(n^3 + \big(n+n^3\big) \epsilon^2r_0^2 \Big).  
\end{align}
Then, the central charge can be estimated as 
\begin{equation}
 c_L = \frac{12 a^2}{G_N}
 \big(1 + \epsilon^2 r_0^2\big) . 
\end{equation}
For the left movers, the temperature is Frolov-Thorne temperature, 
\begin{equation}
 T_L = \frac{1}{2\pi}. 
\end{equation} 
The central charge for right movers is 
calculated in \cite{mty} and the result is 
\begin{equation}
 c_R = \frac{12 a^2 \epsilon}{G_N} . 
\end{equation}
For right movers, the temperature is the Hawking temperature, 
\begin{equation}
 T_R = \frac{r_0}{2\pi} . 
\end{equation}
Then, using the ``Cardy formula'' \eqref{Cardy2}, 
we can obtain the entropy  
\begin{align}
 S &= S_L + S_R 
= \frac{2\pi a^2}{G_N}\Big(1+ \epsilon r_0 + \epsilon^2 r_0^2 \Big) . 
\end{align}
This agrees with the Bekenstein-Hawking entropy $S_{\rm BH}$  
up to $\mathcal O(\epsilon^3)$. 

Before closing,   
we should notice that the higher order corrections in 
the asymptotic Killing vector cannot be fixed in 
the framework of the asymptotic symmetry. 
At the leading and $\mathcal O(\epsilon)$, 
the entropy comes from the leading contributions 
of the left and right movers, respectively. 
However, the $\mathcal O(\epsilon^2)$ term is subleading. 
Hence, the higher order corrections 
of the asymptotic Killing vector can possibly break 
the agreement at $\mathcal O(\epsilon^2)$.

\end{document}